\documentclass[aps,preprint,onecolumn,showpacs,floatfix,superscriptaddress,noshowpacs,longbibliography]{revtex4-1}
\usepackage{graphicx}
\usepackage{amsmath}
\usepackage{amssymb}
\usepackage{color}
\usepackage{float}
\usepackage{hyperref}
\usepackage{natbib}

\begin{document}

\flushbottom

\title{One-dimensional moir\'e charge density wave in the hidden order state of URu$_2$Si$_2$ induced by fracture} 

\author{Edwin Herrera}
\affiliation{Laboratorio de Bajas Temperaturas y Altos Campos Magn\'eticos, Departamento de F\'isica de la Materia Condensada, Instituto Nicol\'as Cabrera and Condensed Matter Physics Center (IFIMAC), Unidad Asociada UAM-CSIC, Universidad Aut\'onoma de Madrid, E-28049 Madrid,
Spain}
\affiliation{Departamento de F\'isica, Universidad Nacional de Colombia, Bogot\'a 111321, Colombia}
\affiliation{Facultad de Ingenier\'ia y Ciencias B\'asicas, Universidad Central, Bogot\'a 110311, Colombia}

\author{V\'ictor Barrena}
\affiliation{Laboratorio de Bajas Temperaturas y Altos Campos Magn\'eticos, Departamento de F\'isica de la Materia Condensada, Instituto Nicol\'as Cabrera and Condensed Matter Physics Center (IFIMAC), Unidad Asociada UAM-CSIC, Universidad Aut\'onoma de Madrid, E-28049 Madrid,
Spain}

\author{Isabel Guillam\'on}
\affiliation{Laboratorio de Bajas Temperaturas y Altos Campos Magn\'eticos, Departamento de F\'isica de la Materia Condensada, Instituto Nicol\'as Cabrera and Condensed Matter Physics Center (IFIMAC), Unidad Asociada UAM-CSIC, Universidad Aut\'onoma de Madrid, E-28049 Madrid,
Spain}

\author{Jos\'e Augusto Galvis}
\affiliation{Facultad de Ingenier\'ia y Ciencias B\'asicas, Universidad Central, Bogot\'a 110311, Colombia}

\author{William J. Herrera}
\affiliation{Departamento de F\'isica, Universidad Nacional de Colombia, Bogot\'a 111321, Colombia}

\author{Jos\'e Castilla}
\affiliation{Laboratorio de Bajas Temperaturas y Altos Campos Magn\'eticos, Departamento de F\'isica de la Materia Condensada, Instituto Nicol\'as Cabrera and Condensed Matter Physics Center (IFIMAC), Unidad Asociada UAM-CSIC, Universidad Aut\'onoma de Madrid, E-28049 Madrid,
Spain}

\author{Dai Aoki}
\affiliation{Universit\'e Grenoble Alpes, CEA, INAC-PHELIQS, 38000 Grenoble, France}

\author{Jacques Flouquet}
\affiliation{Universit\'e Grenoble Alpes, CEA, INAC-PHELIQS, 38000 Grenoble, France}

\author{Hermann Suderow}
\affiliation{Laboratorio de Bajas Temperaturas y Altos Campos Magn\'eticos, Departamento de F\'isica de la Materia Condensada, Instituto Nicol\'as Cabrera and Condensed Matter Physics Center (IFIMAC), Unidad Asociada UAM-CSIC, Universidad Aut\'onoma de Madrid, E-28049 Madrid,
Spain}

\date{\today}

\begin{abstract}
Moir\'e patterns can lead to fundamentally new electronic behavior when formed between two atomic lattices slightly shifted with respect to each other\cite{Bistritzer12233,Cao2018,Xie2019,Choi2019,Jiang2019,Lu2019,doi:10.1021/acsnano.8b08051}. A solid is however not just characterized by the atomic lattice, but also by dynamical charge or magnetic excitations that do not need to be commensurate to the lattice. This raises the question if one can obtain a moir\'e through the interaction of dynamic modes with the atomic lattice. Here we report on the discovery of a one-dimensional charge density wave (1D-CDW) which is a moir\'e pattern between the atomic lattice and a hot spot for electronic scattering in the bandstructure of the hidden order (HO) state of URu$_2$Si$_2$. The moir\'e is produced by fracturing the crystal in presence of a dynamical spin mode at low temperatures. Our results suggest that charge interactions are among the most relevant features competing with HO in URu$_2$Si$_2$.
\end{abstract}

\maketitle

The HO state of URu$_2$Si$_2$ puzzles experimentalists and theoreticians alike since its discovery more than three decades ago \cite{FLOUQUET2005139,Mydosh11,Mydosh_2020}. It consists of a phase transition below $T_{HO}=17.5$ K in which there is a large entropy change\cite{Wang_2020}. However, neutron scattering experiments do not show any sign of static magnetic nor structural order\cite{doi:10.1080/14786435.2014.935513}. Instead, there are dynamical modes of Ising-like magnetic excitations with a strong fluctuating magnetic moment, $\mu\approx 1-2\mu_B$, located at $q_0=(0,0,1)$ and $q_1=(0.6,0,0)$\cite{Broholm87,Broholm91,doi:10.1080/14786435.2014.935513}. These modes are quenched into an antiferromagnet order above 5 kbar with the $q_0$ wavevector\cite{doi:10.1080/14786435.2014.935513,Mydosh_2020,Villaume08} and with the $q_1$ wavevector at high magnetic fields\cite{Wiebe07,Villaume08,Knafo16}. The magnetic field also modifies the Fermi surface, producing nesting at $q_1$\cite{doi:10.7566/JPSJ.82.034706}. In addition, several measurements indicate that at zero field the bandstructure has a hot spot for scattering at $q_1$, so that there is a strong interaction between electronic degrees of freedom and the dynamical modes\cite{Elgazzar09,doi:10.7566/JPSJ.82.034706,Wiebe07,Morr_2016}. Here we use Scanning Tunneling Microscopy (STM) to study high quality single crystals of URu$_2$Si$_2$. Contrasting previous STM work\cite{Aynajian10, Schmidt10}, we find that there is a one-dimensional charge modulation with a wavevector that is a moir\'e combination of the atomic lattice periodicity and the dynamical spin excitation mode at $q_1$. This suggests that the dynamical modes can lead to new symmetry breaking ground states in URu$_2$Si$_2$ through external action.

\begin{figure}
	\begin{center}
	\centering
	\includegraphics[width=0.95\textwidth]{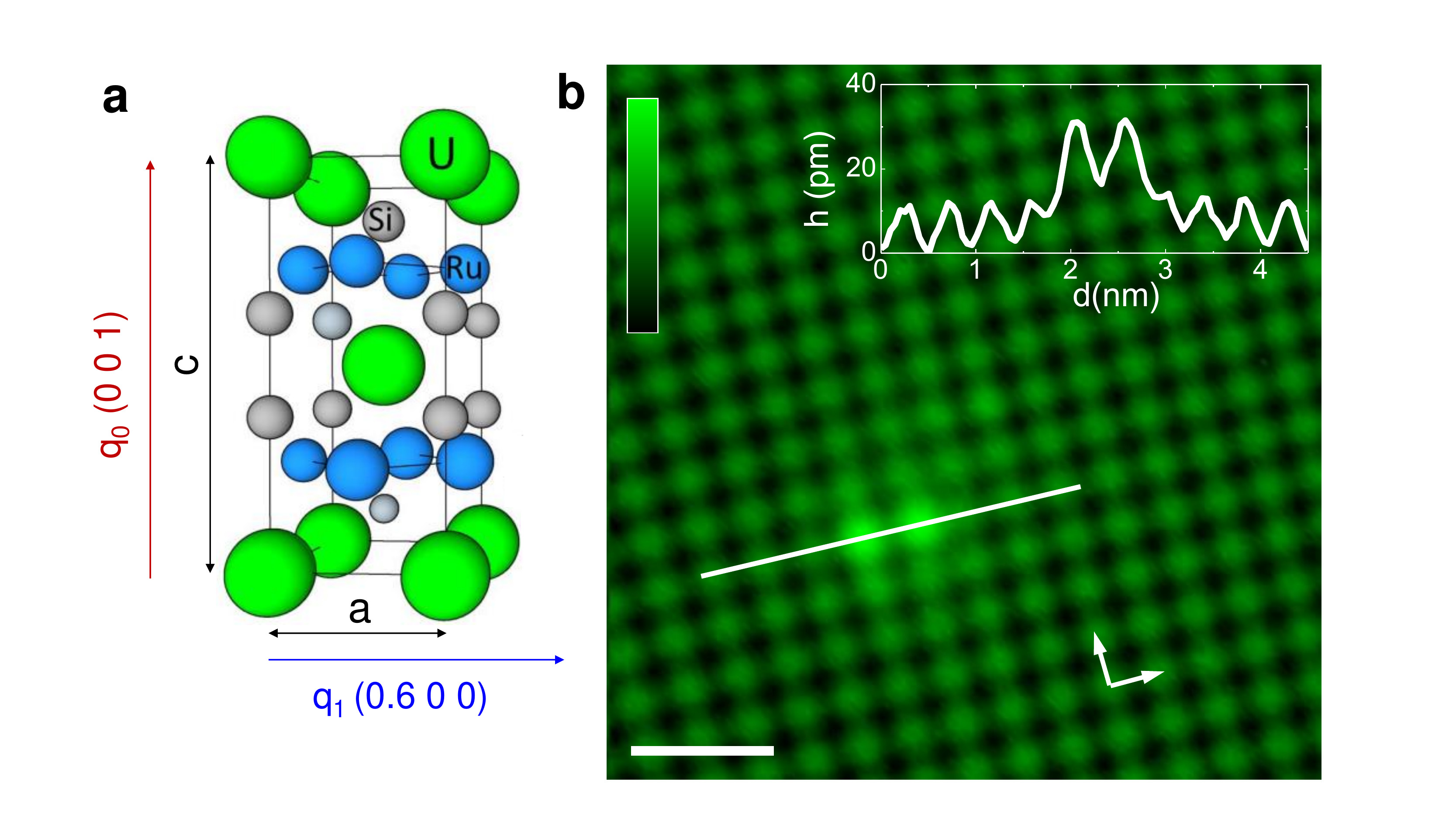}
	\end{center}
	\vskip -0.5 cm
	\caption{{\bf Atomic size STM images of the square in-plane lattice in URu$_2$Si$_2$.} {\bf a} URu$_2$Si$_2$ tetragonal crystalline structure ($a=420$\,pm\,and $c=960$\,pm), with U atoms in green, Ru atoms in blue and Si atoms in grey. The directions of the propagating vectors of the dynamic magnetic excitations of the HO state are schematically shown by red ($q_0$ mode along the c-axis) and blue ($q_1$ mode in-plane) arrows. {\bf b} Atomic resolution STM image showing the square U surface. A defect consisting of a couple of atomic size features is seen in the image. The white line provides the path for the scan shown in the inset (upper right). The bias voltage is of 5.5 mV and the current is held constant at 3 nA. White arrows show the in-plane crystalline directions and the color scale the height difference, following the color bar on the upper left (from zero, black to 40 pm, green). White scale bar is 1.3 nm long.}
	\label{Fig1_Structure}
	\end{figure}

	\begin{figure}
	\begin{center}
	\centering
	\includegraphics[width=0.65\textwidth]{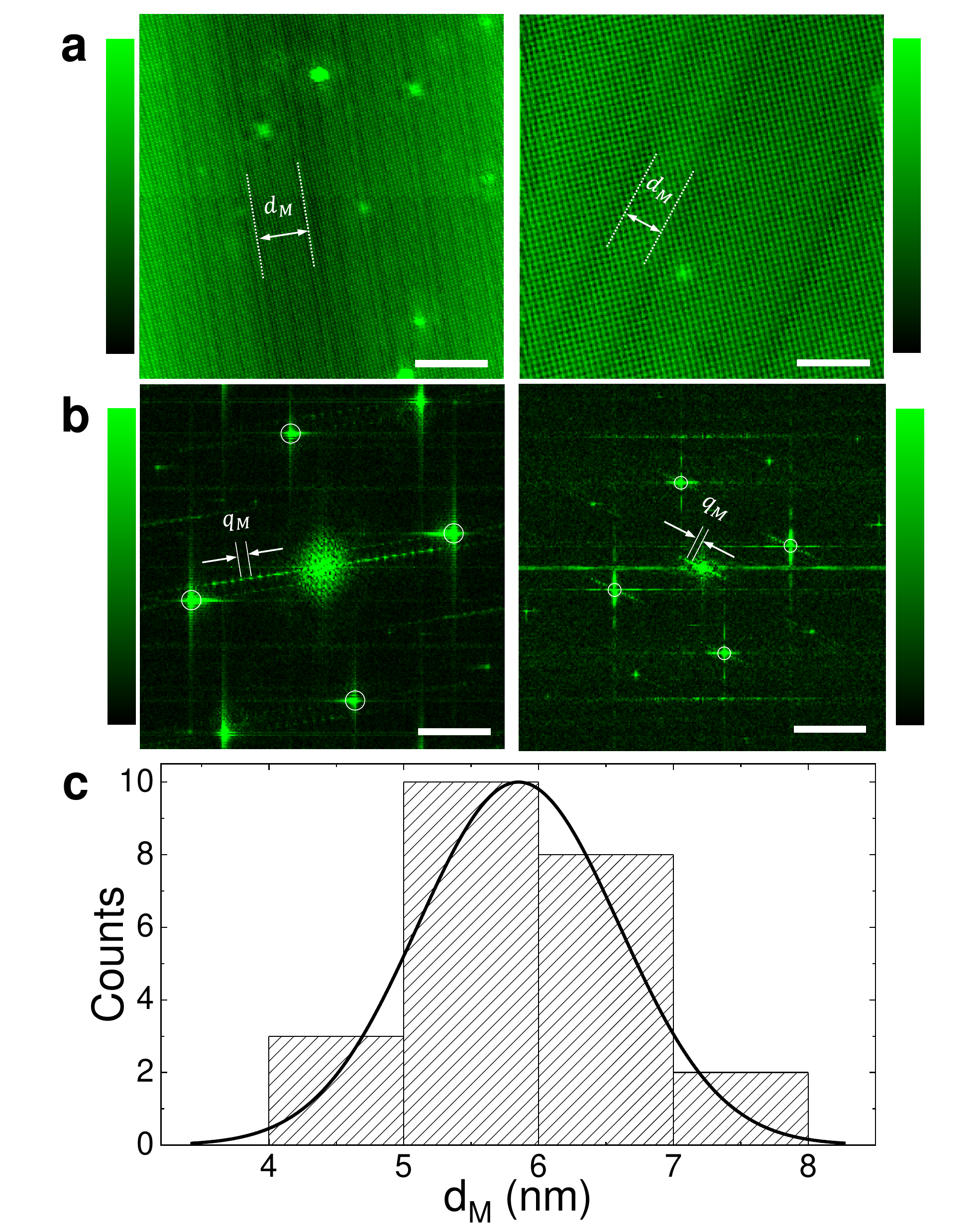}
	\end{center}
	\vskip -0.5 cm
	\caption{{\bf Moir\'e 1D-CDW in URu$_2$Si$_2$.} {\bf a} Atomic resolution STM images in two different fields of view. Notice that these are considerably larger than those in Fig.\,\ref{Fig1_Structure}. We mark the one-dimensional modulation of wavelength $d_M$ by dashed lines and the arrow. The current is constant at 0.5 nA and the bias voltage is of 4 mV in left panel and 0.05 nA and 5 mV in the right panel. The colored bars provide the vertical scale, with 20 pm from black to green. Horizontal scale bars are 8.4 nm (left panel) and 4.7 nm (right panel) long. {\bf b} Fourier transform of the images. We observe the usual atomic Bragg peaks due to the square surface atomic lattice and also multiple peaks at reciprocal space distances $q_M$. We mark with white circles the Bragg peaks of the atomic lattice and the wavevector of the modulation $q_M$. Horizontal scale bars are 1.2 nm$^{-1}$ (left panel) and 2 nm$^{-1}$ (right panel) long. The Fourier amplitude is given by the vertical bars, with 0.6 pm in left panel and 1.6 pm in right panel, from black to green. In {\bf c} we show a histogram over $d_M$ obtained in 24 different fields of view acquired in 5 different samples. We provide images made in all samples and their pictures after the cleaving process in the Supplementary Information.}
	\label{Fig2_Modulation}
	\end{figure}

	\begin{figure*}
	\begin{center}
	\centering
	\includegraphics[width=\textwidth]{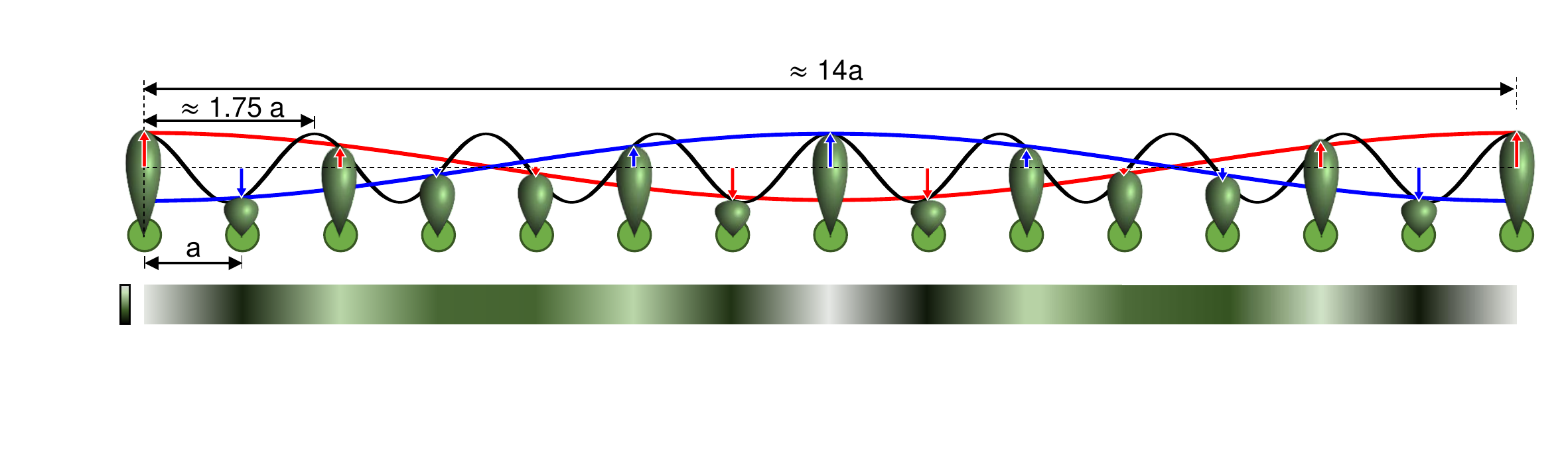}
	\end{center}
	\vskip -0.5 cm
	\caption{{\bf Schematic of the moir\'e 1D-CDW.} We represent the atoms in green in form of a single orbital, whose elongation in z is coupled to a modulation located at the reciprocal lattice vector $q_1=0.57\approx 0.6$ (black line, $\lambda_1 \approx 1.75a$ in real space, with $a=420$ pm). The elongation is highlighted by a colored arrow at each atomic position. We represent the resulting charge modulation using a color scale by the bar at the bottom. The wavevector of the modulation is $q_M$ ($\lambda_M\approx 14a$ in real space). Because $\lambda_1/2\approx a$, the relation between $\lambda_1$ and $a$ is through a "fractional" moir\'e $\lambda_M=\frac{\lambda_1\cdot a}{| \lambda_1/n - a|}$ with $n=2$ and the pattern is formed by two cosine waves with the same wavelength, but shifted by $\pi$ from each other (red and blue).}
	\label{Fig3_ModulationModel}
	\end{figure*}

Figure \ref{Fig1_Structure}(a) shows the crystal structure of URu$_2$Si$_2$ together with a representation of the directions of $q_0$ and $q_1$. In Fig.\,\ref{Fig1_Structure}(b) we show atomically resolved images of the U square atomic lattice, with the in-plane lattice parameter  $a=420 $pm, similar as in previous work \cite{Aynajian10, Schmidt10,PhysRevB.98.115121}. We focus here mostly on features obtained from the STM topography. Spectroscopic features are mentioned in the supplementary information and mostly coincide with previous work.

	\begin{figure}
	\begin{center}
	\centering
	\includegraphics[width=0.95\textwidth]{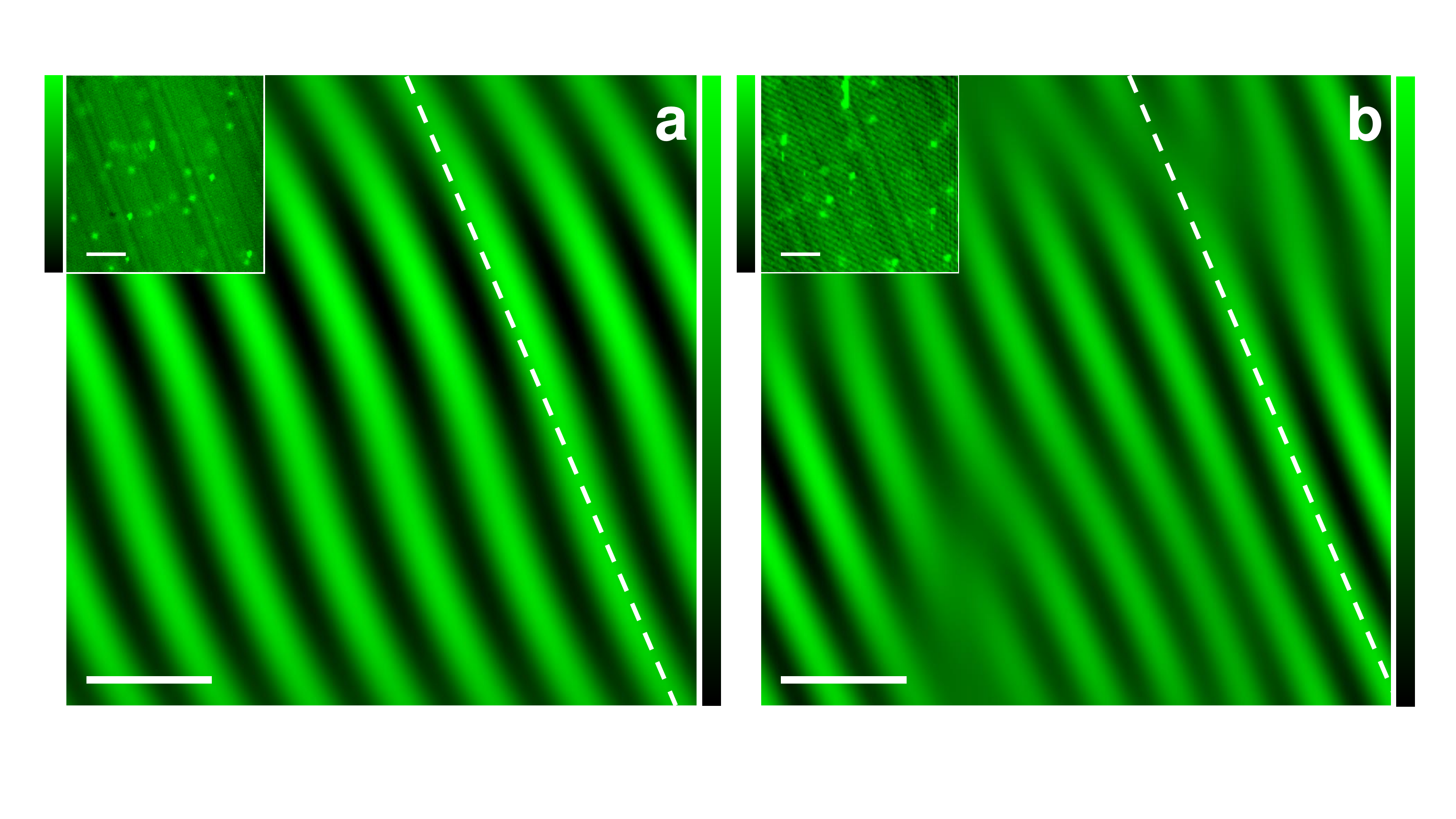}
	\end{center}
	\vskip -0.5 cm
	\caption{{\bf Defects in the moir\'e 1D-CDW.} In {\bf a} and {\bf b} we show two fields of view where we have Fourier filtered the first Bragg peaks of the modulation. The corresponding images are shown in the insets (current of 1.4 nA and bias voltage of 2 mV in both images, vertical scale bar is shown in the left, with 30 pm in the inset of {\bf a} and 20 in the inset of {\bf b} from black to green). White dashed lines are guides to the eye following linear features. Fourier filter is made exactly at two Bragg peaks of the one-dimensional modulation. Color scale is given on the right, with 2 pm from black to green. White scale bars are 10 nm long in all images.}
	\label{Fig4_phaseshift}
	\end{figure}

In fields of view that are sufficiently large and {\color{blue}free of steps}, there is a certain pattern which repeats on the image and is one-dimensional (Fig.\ref{Fig2_Modulation}). The pattern corresponds to height changes which vary from place to place and are very small, of at most 2 pm,i.e. 1\% or less as compared to the usual atomic corrugation (inset of Fig.\,\ref{Fig1_Structure}{\bf b}, \cite{Aynajian10, Schmidt10,PhysRevB.98.115121}). The period and the direction of the one-dimensional pattern can be determined from the Fourier transforms of the topographic images. The Bragg peaks associated to the one-dimensional modulation repeat at integer multiples of $q_M$. We determine $q_M$  from the reciprocal space distance between consecutive peaks and obtain that $q_M  = 0.17 1/$nm (Fig.\ref{Fig2_Modulation}{\bf b}).  This corresponds to a modulation of period $d_M = 1/q_M = 6$\,nm ($d_M=14a$). The one-dimensional modulation is nearly parallel to an in-plane crystalline direction, often with a small angle of a few degrees. A histogram with observed $d_M$ is shown in Fig.\,\ref{Fig2_Modulation}{\bf c}. We observed the modulation in many fields of view on five different samples, each one cleaved at low temperatures (more details in the Supplementary Information), up to magnetic fields of 4 T and temperatures several K above liquid helium. At higher temperatures and above T$_{HO}$ we were unable to detect it because the amplitude of the modulation is too small to detect in presence of temperature induced drift in the STM. As discussed in the Supplementary Information in detail, the observed features are associated to fracturing the sample at low temperatures.

A modulation observed in STM topographic images superimposed to the atomic lattice is usually caused by charge order in form of a CDW. The Fermi surface has mainly a four-fold symmetry and there are no features in the Fermi surface which could lead to any 1D-CDW\cite{Elgazzar09,Bourdarot10}. Features of the Fermi surface are instead fourfold, showing nesting at $q_0$ along the c-axis. Nesting occurs in-plane at $q_1$ too at sufficiently high magnetic fields\cite{doi:10.7566/JPSJ.82.034706}. At the level of the local atomic size density of states measured by in-plane topographic imaging with STM, we do not expect to observe directly modulations at wavevectors corresponding to dynamical modes. However, we might be sensitive to effects related to the mode at $q_1$, which is clearly an in-plane hot spot in the Fermi surface. As we show in the following, $q_M$ is related to $q_1$ through a moir\'e pattern and the direction of $q_M$ is determined by the direction of the propagation of the crack front during fracture.

First, let us establish the relation between $q_1$ and $q_M$, knowing that $q_1 \gg  q_M$. The superposition of 1D modulations with similar periodicities $\lambda_1$ and $\lambda_2$ ($\lambda_1\approx\lambda_2$) leads to an additional modulation, the moir\'e pattern, at a scale which is far above $\lambda_1$ and $\lambda_2$ and is given by $\lambda_M=\frac{\lambda_1\cdot \lambda_2}{| \lambda_1 - \lambda_2|}$\cite{Zhou:08,10.2307/24936147}. If $\lambda_2=a$ and $\lambda_1\approx na$ with $n$ an integer (and of course $\lambda_1\ne na$), the moir\'e adopts a "fractional" form and $\lambda_M=\frac{\lambda_1\cdot a}{| \lambda_1/n - a|}$. The moir\'e is then composed of $n$ cosine functions of wavelength $\lambda_M$ phase-shifted to each other by $2\pi/n$. In STM images made at constant tunneling current, we measure the variations in the local density of states (LDOS) integrated between the Fermi level and the bias voltage as a function of the position. The LDOS can be understood as a combination of localized orbitals and itinerant electrons\cite{PhysRevLett.50.1998,PhysRevB.31.805}. Let us take for simplicity the case of a 1D row of atoms, separated by a given lattice constant. If there is a very weak modulation at a wavevector close to an integer times the lattice constant superposed to the 1D atomic lattice, the visible signature of this modulation on the LDOS is formed by the moir\'e pattern resulting from the value of the modulation at the atomic lattice positions. Let us consider schematically a set of elongated atomic orbitals located at integer multiples of $a$ and assume that the size of the lobes is given by a modulation close to $q_1$ (at $\approx$ (0.57 0 0), which gives $\lambda_1\approx 2\lambda_2$, with $\lambda_1$ the modulation at $q_1$, $\lambda_1=1/q_1$, and $\lambda_2=a$). The result is schematically shown in Fig.\,\ref{Fig3_ModulationModel}. The size of the lobes leads to a pattern which consists of two inverted cosines, each one with a periodicity $\lambda_M=\frac{\lambda_1\cdot a}{| \lambda_1/2 - a|}$ and shifted by $\pi$. Thus, our images show CDW at $\lambda_M$ as a consequence of a moir\'e combination between $q_1$ and the atomic lattice.

Second, to understand the symmetry breaking, let us discuss the origin of the modulation. The modulation is not present in any of the STM images shown in Ref.\cite{Aynajian10, Schmidt10}, although their resolution was of the same order as ours. In those experiments, samples were cleaved at relatively high temperatures (of the order of liquid nitrogen), whereas we cleave our sample at very low temperatures (below liquid helium). Cleavage of a hard single crystal is equivalent to brittle fracture. It is a fast procedure by which the bonds are broken first on one side of the sample and then a crack front travels through the sample in a short amount of time. The propagation of the crack front during cleavage leaves permanent modifications of the surface\cite{PhysRevE.67.066209}. These modifications can be due to acoustic waves that are emitted when a crack front crosses defects during the cleaving process or due to other forms of local interaction of the crack front and the crystalline lattice and have been extensively studied in materiales such as Si, sapphire or tungsten\cite{SHERMAN20041743,PhysRevE.67.066209,Kermode2013,Zhao2018}. From a careful optical and electron microscopy analysis of surfaces of URu$_2$Si$_2$ broken at room temperatures and at low temperatures (see Supplementary Information) we obtain that samples cleaved at low temperatures have a set of linear features created during crack and oriented close to a crystalline axis. These features define a fixed direction which is close to one of the main in-plane crystalline axis and produce the one-dimensional symmetry breaking field. We note that we observe the modulation in five samples cut differently, each one with slightly different shapes, different crack initiation points, and different internal structure of defects.

The modulation is not perfect over the whole crystal but shows defects at a few locations. Such a situation occurs in Fig.\,\ref{Fig4_phaseshift}, where we filter the peaks corresponding to the modulation at $q_M$ to obtain Fig.\,\ref{Fig4_phaseshift}{\bf a,b}. We observe that, in these fields of view, the stripes are structured. In  Fig.\,\ref{Fig4_phaseshift}{\bf a} there is a shear shift of the charge modulation. In  Fig.\,\ref{Fig4_phaseshift}{\bf b} there is a pair of dislocations in the charge modulation in the top of the image and another dislocation in the bottom of the image.

Our scans show the 1D-CDW Bragg peaks just at the center of the Fourier transform. To understand this observation, let us describe a CDW through a vector $\boldsymbol{D(\boldsymbol{r})}=\boldsymbol{A}cos(\boldsymbol{q}\boldsymbol{r}+\phi)$, with $\boldsymbol{A}$ the vector of the absolute value of the modulation, $\boldsymbol{q}$ the wavevector of the modulation and $\phi$ a phase shift\cite{ElBaggari1445}. As we show in the Supplementary Information, the corresponding Fourier transform shows satellite peaks around lattice Bragg peaks when the displacement vector is in-plane along $\boldsymbol{A}=(A_0,0,0)$ or $\boldsymbol{A}=(0,A_0,0)$. If the displacement is out-of-plane, $\boldsymbol{A}=(0,0,A_0)$, the image shows satellite peaks around the center of the Fourier transform. The latter is always observed in our images. Thus, we are mostly observing an amplitude modulation of the charge (out-of-plane), without in-plane atomic displacements.

Notice that there is no in-plane strain associated to the modulation, although the modulation might influence strain caused by defects. To analyze this possibility, we have calculated the strain from our images following Refs.\,\cite{Lawler2010,Zeljkovic2015}. We can identify slight modifications of the strain built up around defects consisting of Si atoms on top of a U surface, with a tendency to form strain with a slightly elongated shape. It is useful to remind that Nuclear Magnetic Resonance (NMR) measurements highlight the relevance of charge disturbances and in particular show a two-fold lineshape on Si associated to defects\cite{Kambe13,PhysRevB.91.035111,PhysRevB.97.235142}. Our results suggest that Si defects are influenced by charge disturbances such as the one-dimensional charge modulation.

We should note that crack front velocity is of order of sound velocity (see Supplementary Information and\cite{SHERMAN20041743,PhysRevE.67.066209,Kermode2013,Zhao2018}). Thus, we can link together the external action (crack front), atomic displacement modes (phonons) and the electronic susceptibility of URu$_2$Si$_2$ (fluctuations at $q_1$). The phonon dispersion in URu$_2$Si$_2$ shows no strong features connected to the HO transition\cite{PhysRevB.91.035129}. However, there is an anomalous phonon mode broadening which suggests strong anharmonicity and coupling to magnetic excitations\cite{PhysRevB.91.035128,PhysRevB.91.035129,PhysRevB.93.075123}. The dispersion relation of the $q_1$ modes provides a velocity $v=\frac{d\epsilon}{dk}\approx 10^{4}\frac{m}{s}$\cite{PhysRevB.91.035129}. Most remarkably, crack front velocity, sound velocity and the velocity of the magnetic $q_1$ modes are all of the order of tens of km/s. 

Interestingly, the time required for each of these three modes to cross a single unit cell is in the femtosecond regime. Thus, ultrafast radiation experiments on surfaces of URu$_2$Si$_2$ might lead to similar effects as a travelling crack front and potentially to the nucleation of a similar CDW.

Notice that moir\'e patterns arise on surfaces or few layer systems because of displacements or rotations of atomic lattices\cite{Bistritzer12233,Cao2018,Xie2019,Choi2019,Jiang2019,Lu2019,doi:10.1021/acsnano.8b08051}. Rotations among atomic layers provide a control parameter, the relative angle between layers, to modify the moir\'e\cite{Cao2018}. In our case, moir\'e modulations can arise with any characteristic vector from the electronic bandstructure. That is, with wavevectors located inside the Brillouin zone and lying close to an integer fraction of the unit cell lattice. To modify the moir\'e, we need to modify the bandstructure, with the usual control parameters as doping, stress or magnetic field. For very large values of $n$ in $\lambda_M=\frac{\lambda_1\cdot a}{| \lambda_1/n - a|}$, the moir\'e pattern might rapidly acquire a very long wavelength, which should make it quite difficult to distinguish from a large defect or distortion. However, for small $n$ (and with a sufficient difference between $\lambda_1$ and $na$), the moir\'e combinations proposed here could lead to a variety of new ground states. 

In summary, we observe that the high susceptibility of URu$_2$Si$_2$ at $q_1$ results in quenched 1D-CDW when there is a sufficiently strong interaction with a cracking process. But we are not just condensing a CDW related to $q_1$ through fracture. The moir\'e between the atomic lattice and the $q_1$ modulation leads to a real, physical, 1D-CDW which breaks the in-plane square C4 symmetry. Of course such a CDW is not a property of the HO. But it is related to HO through one of the wavevectors that lead to the moir\'e. Our observation supports the claim that in the HO phase electric interactions might play a key role, as proposed by nuclear magnetic resonance experiments\cite{PhysRevB.97.235142}.

\section*{Acknowledgments}

We acknowledge discussions with F. Guinea, A. Levy Yeyati and with S. Vieira and support by the Spanish MINECO (FIS2014-54498-R, MDM-2014-0377), by the Comunidad de Madrid through program NANOMAGCOST-CM (Grant No. S2018/NMT-4321) and by Cost Action CA16218 (Nanocohybri). I.G. acknowledges support by the European Research Council PNICTEYES grant agreement 679080. We also acknowledge  the support of Departamento Administrativo  de Ciencia, Tecnolog\'ia  e  Innovaci\'on,  COLCIENCIAS  (Colombia) Convocatoria 784 - 2017 and the Cluster de investigaci\'on en ciencias y tecnolog\'ias convergentes de la Universidad Central (Colombia).  We also acknowledge SEGAINVEX at UAM.

\section*{Author contributions}
E.H. and I.G. spotted the one-dimensional modulation and worked through its characterization. The experiment was proposed by D.A. and J.F. and designed by E.H., I.G. and H.S.. V.B. searched for possible combinations of modulations, identified the moir\'e and calculated the strain, with the supervision of H.S., E.H. and I.G.. W.J.H. and J.A.G. provided support to the interpretation and in analyzing fracture, together with E.H. and with input from H.S.. J.C. made the corresponding simulations. D.A. prepared and characterized samples. The paper was written by E.H., I.G., J.F. and H.S. with contributions from all authors.

\section*{Methods}

We  use a STM set-up  described in Refs.\cite{Suderow11,Galvis15}  that  features a movable sample holder which we use to cleave in-situ the sample and change in-situ the scanning window\cite{Suderow11}. The single crystals of URu$_2$Si$_2$ have been grown by Czochralski method and have a residual resistance ratio of about 120 \cite{Aoki10}. We first screened crystals for high quality, from different growths and then cut crystals into needles of about 0.5 mm $\times$ 0.5 mm $\times$ 2 mm. We inserted each needle into a hole made on a gold substrate and glued them with silver epoxy. The needles were positioned into the sample holder in such a way that they hit a sharp ceramic blade when pulling on the sample holder. Slightly below 4.2 K we pulled on the holder through a manual mechanism located at room temperature and connected to the holder with a piano cord. We felt resistance when the sample touched the blade. We continued pulling until we heard a characteristic "crack". At this point, the sample was broken. The noise created during crack, which we heard outside the dilution refrigerator, evidences generation of sound waves during fracture. This was made with the tip far from the sample. We then approached the tip to the sample, scanned and obtained the results discussed here. We provide results obtained in five succesful cleaves, obtaining each time atomically flat surfaces in cryogenic vacuum. In each crystal, we were able to study several tens of scanning windows, each limited by the scanning range of our piezotube (2$\mu$m $\times$ 2$\mu$m)\cite{Suderow11}. We present almost exclusively topographic STM images made at 150 mK and with the STM on constant current mode held by a feedback loop, with a setpoint of a fraction of nA and at a bias voltage of order of 10 mV.

\newpage

\section*{Supplementary Information}

\subsection{Further details of the experimental set-up and results on five samples}

\begin{figure}
	\begin{center}
	\centering
	\includegraphics[width=0.9\textwidth]{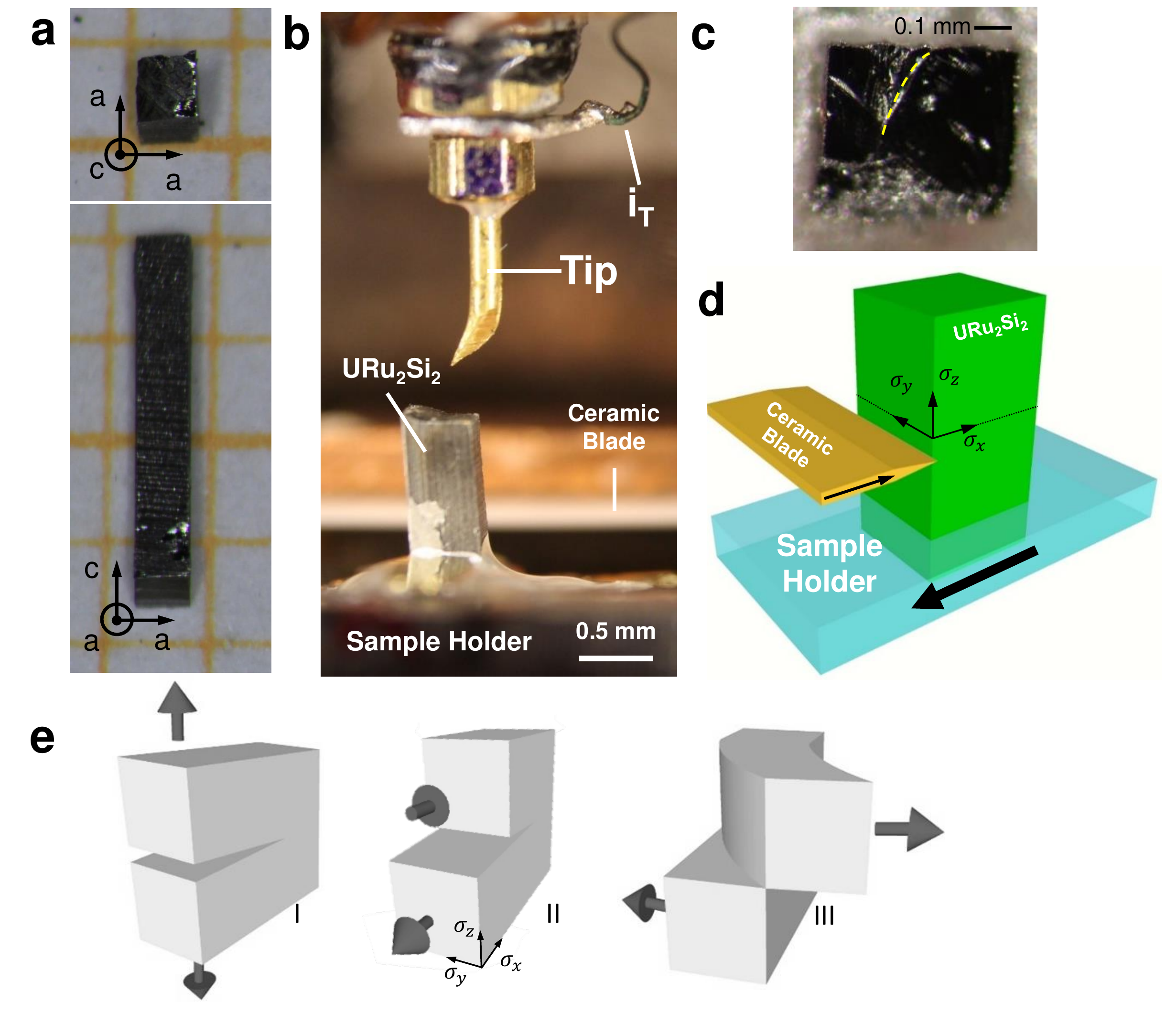}
	\end{center}
	\vskip -0.5 cm
	\caption{{\bf Method used to cleave the URu$_2$Si$_2$ single crystals.} In {\bf a} we show a sample of URu$_2$Si$_2$ from the side
(bottom) and from the top (top) on top of a mm  scale grid. We mark the crystalline
axes with arrows. The sample has been cut using a wire saw.
In {\bf b} we show the sample mounted on the STM at room
temperature. The sample is glued using silver epoxy on the
sample holder. The tip of Au is positioned on the top of the
sample. After cleaving at liquid helium temperatures, the tip
moves using a piezoelectric motor down to tunnel into the
broken sample (not shown). We also mark the wire for the
tunneling current. The contact to the sample is made through
the sample holder. The ceramic blade is also seen. It is located
(out of focus in the picture) behind the sample. In {\bf c}
we show a photography of a sample broken at low temperatures,
surrounded by the silver epoxy. The yellow dashed line is a large defect inside the crystal. At low temperatures,
the sample holder is moved towards the blade, as shown in
{\bf d}. The efforts made on the sample are shown in {\bf e} and consist
of tear (mode I), compression (mode II) and shear (mode
III). In {\bf d,e} we mark the corresponding stress $\sigma_{x,y,z}$ as black
arrows.}
	\label{Fig5_Setup}
	\end{figure}

The STM set-up has been described in detail elsewhere\cite{Suderow11}. However, it is useful to provide details about the additions that we have made to be able to cleave at low temperatures hard samples as URu$_2$Si$_2$. In Fig.\,\ref{Fig5_Setup}{\bf a} we show pictures of URu$_2$Si$_2$ from the side and from the top. The sample has been shaped using a wire saw into a rectangle elongated perpendicular to the cleaving plane (the (a,a) axis). In Fig.\,\ref{Fig5_Setup}{\bf b} we show a picture of the sample mounted on the STM. The picture is taken from the front. The piezotube is located on the top of the picture. We can see the tip of Au on the top of the image and the part of the  URu$_2$Si$_2$ that holds out from the sample holder. A cleaved surface is shown in Fig.\,\ref{Fig5_Setup}{\bf c} from the top. In Fig.\,\ref{Fig5_Setup}{\bf d} we schematically show the process of cleaving. We move laterally the sample holder and a ceramic blade pushes laterally the rectangular sample until it breaks. The efforts during the crack are tear, compressive and shear efforts, as schematically represented in Fig.\,\ref{Fig5_Setup}{\bf e}.

The tip of Au is prepared and shaped as shown in Refs.\,\cite{Rodrigo04,Suderow11}. To this end, we glue a pad of Au on the sample holder in such a way that we can move the tip between sample and pad of gold, once the sample has been cleaved. To analyze the images we use WSxM \cite{doi:10.1063/1.2432410} and software available at \cite{Software}.

\begin{figure*}
	\begin{center}
	\centering
	\includegraphics[width=0.85\textwidth]{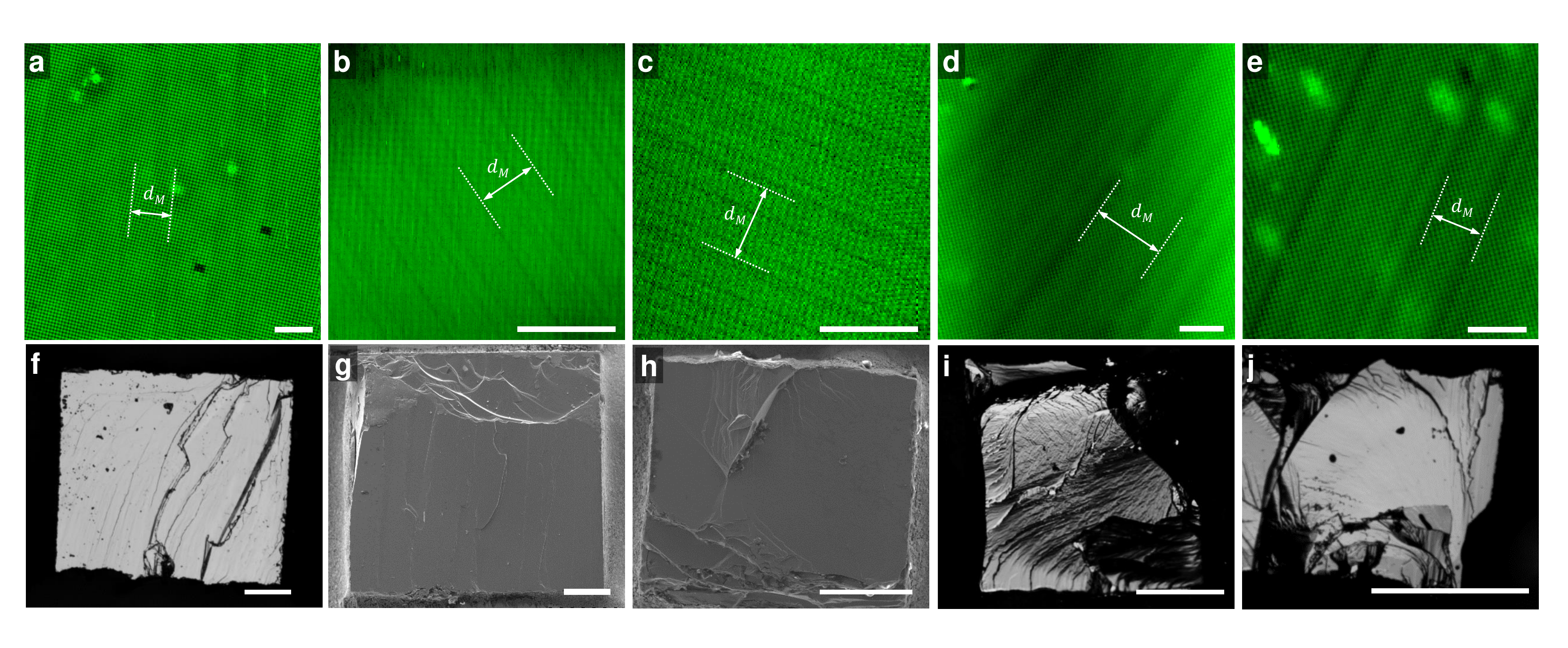}
	\end{center}
	\vskip -0.5 cm
	\caption{{\bf Moir\'e one-dimensional charge density wave on five different samples.} {\bf a-e} Topographic STM images made in different samples and different cleaves at low temperatures. We mark the one-dimensional moir\'e modulation by dashed white lines. The arrows provide the distance obtained from the Fourier transform as described in the text ($d_M$). White bars are 5 nm long. Notice that all are images taken in between large steps ({\bf a} current of 3 nA and bias voltage of 5.5 mV, {\bf b} current of 9 nA and bias voltage of 50 mV, {\bf c} current of 4.8 nA and bias voltage of 5 mV, {\bf d} current of 0.1 nA and bias voltage of 2 mV, {\bf e} current of 1.7 nA and bias voltage of 2 mV). {\bf f-j} Corresponding SEM or optical images of the samples after low temperature cleaving. White scale bars are 200 $\mu$m long.}
	\label{Fig6_Samples}
	\end{figure*}

In Fig.\ref{Fig6_Samples} we show results obtained in five different samples that were cleaved at low temperatures. We see that we can observe the one-dimensional modulation in each sample. The resulting cleaved surfaces are varied. Optical inspection mostly shows large mirror shiny surfaces (Fig.\ref{Fig6_Samples}{\bf f,g,h}) or mirror like surfaces interspersed with large cracks (Fig.\ref{Fig6_Samples}{\bf i,j}). In spite of that we could always obtain atomic resolution and the one-dimensional modulation.

	\begin{figure}[h!]
	\begin{center}
	\centering
	\includegraphics[width=0.65\textwidth]{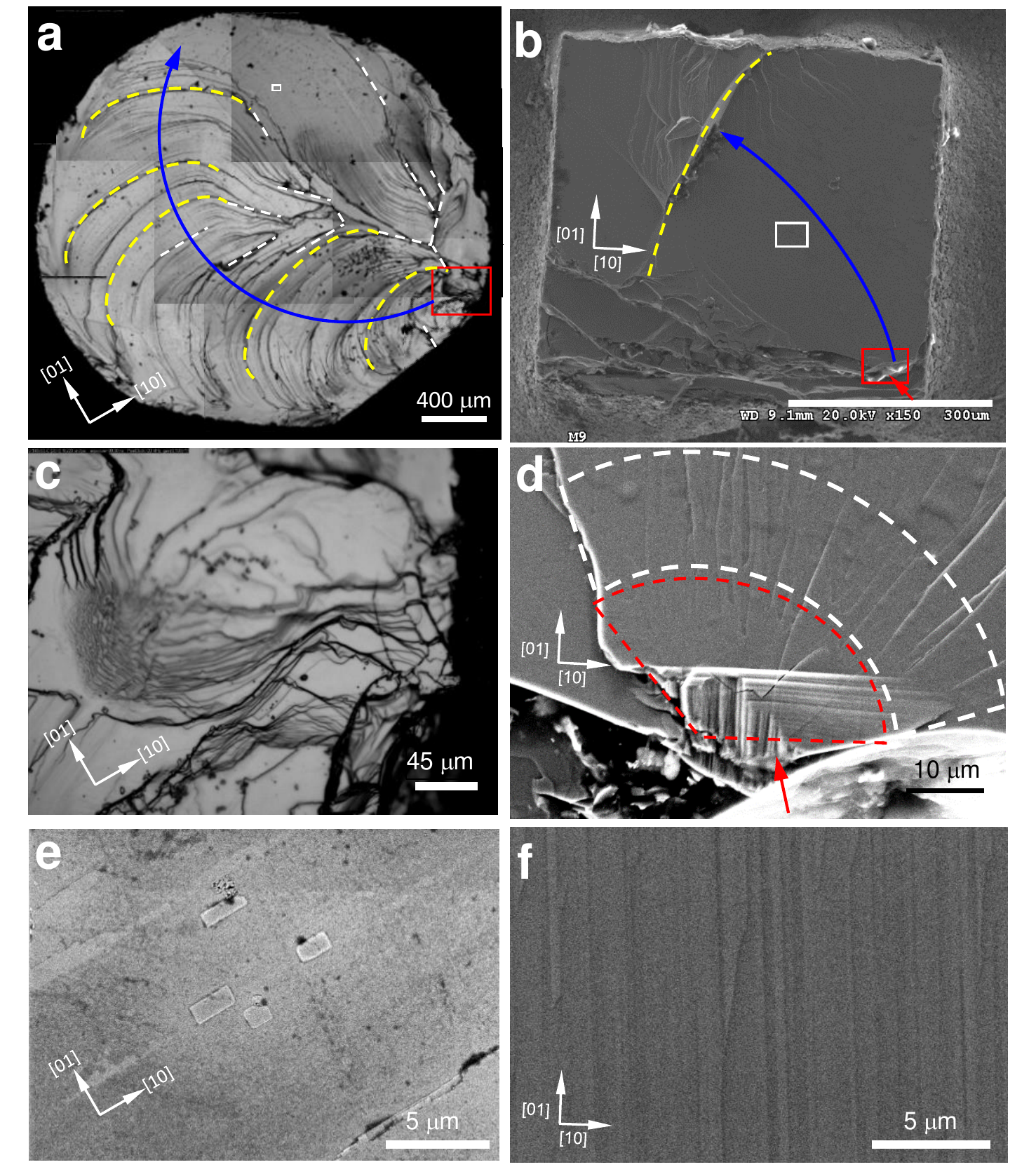}
	\end{center}
	\vskip -0.5 cm
	\caption{{\bf Analysis of surface marks after cleaving.} In {\bf a,c,e} we show pictures of a sample cleaved at room temperature and in {\bf b,d,f} of a sample cleaved at liquid helium. In {\bf a} we show a large scale optical image of the sample. We mark by a blue arrow the path of the crack front. Yellow lines mark remarkable features formed during crack. With a red rectangle we mark the place where the crack started. An optical image of the red rectancle of the room temperature cleaved sample is shown in {\bf c}. A zoom of the region marked by a small white rectangle in {\bf a} is shown in {\bf e}. Notice the presence of defects, but of no well-defined pattern. In {\bf b} we show a SEM image of the sample cleaved at liquid helium temperatures. We mark the region where the crack started by a red rectangle. A zoom of this area is shown in the SEM image {\bf d}. A red arrow identifies the point where the crack started. The dashed red region is the so-called mirror zone, which appears prior to the zone where the crack produces twist hackles distributed radially (white dashed lines). These then proliferate through the sample and, when leaving the mirror zone, tend to align to the crystalline axis. In {\bf f} we show the region on a white rectangle of {\bf b}. We can identify a strongly one-dimensional pattern, which is absent in the other sample ({\bf e}). Arrows in each image provide the crystalline directions. Scale bars are shown in white.}
	\label{Fig7_SEM}
	\end{figure}

\subsection{Analysis of fractured surfaces}

After the experiment, we verified that the only source for the cracking noise was the sample (no components of the dilution refrigerator or the STM were damaged). The remaining part of the sample generally flew away to the bottom of the vacuum chamber. We collected both parts of the sample and made a detailed analysis of their surface using Scanning Electron Microscopy (SEM) and optical microscopy at room temperature.

In single crystals, fractures are the result of the external applied stress, the elastic properties of the crystal and the relative orientation between the crack propagation and the crystalline lattice directions. As a result, characteristic surface marks can be identified to reconstruct the fracture process. We have made a microscopy study of surfaces on URu$_2$Si$_2$ obtained after cleaving at room temperature and at low temperatures (Fig.\,\ref{Fig7_SEM}). We identify with yellow dashed lines step hackles. In both cases, hackles follow an elliptical shape. The crack front travels along a direction perpendicular to the hackle lines\cite{Kermode2013,Zhao2018,Sherman2003}. The curved crack direction results due to the deflection of the crack between planes because of the anisotropic velocity depencence of lattice vibrations\cite{SHERMAN20041743}. In our case, the crack front in the sample cleaved at room temperature can be approximated by an ellipse of semi-axes of $a \approx 2.06 mm$ and $b \approx 2.73 mm$. From such a shape we can estimate the propagation velocity of the crack front to be of order of $10^3$ $m/s$\cite{Kermode2013,Zhao2018,Sherman2003}. The sample cleaved at low temperatures is too small to make an estimation, but we can expect similar crack front velocities.

\begin{figure*}
	\begin{center}
	\centering
	\includegraphics[width=0.95\textwidth]{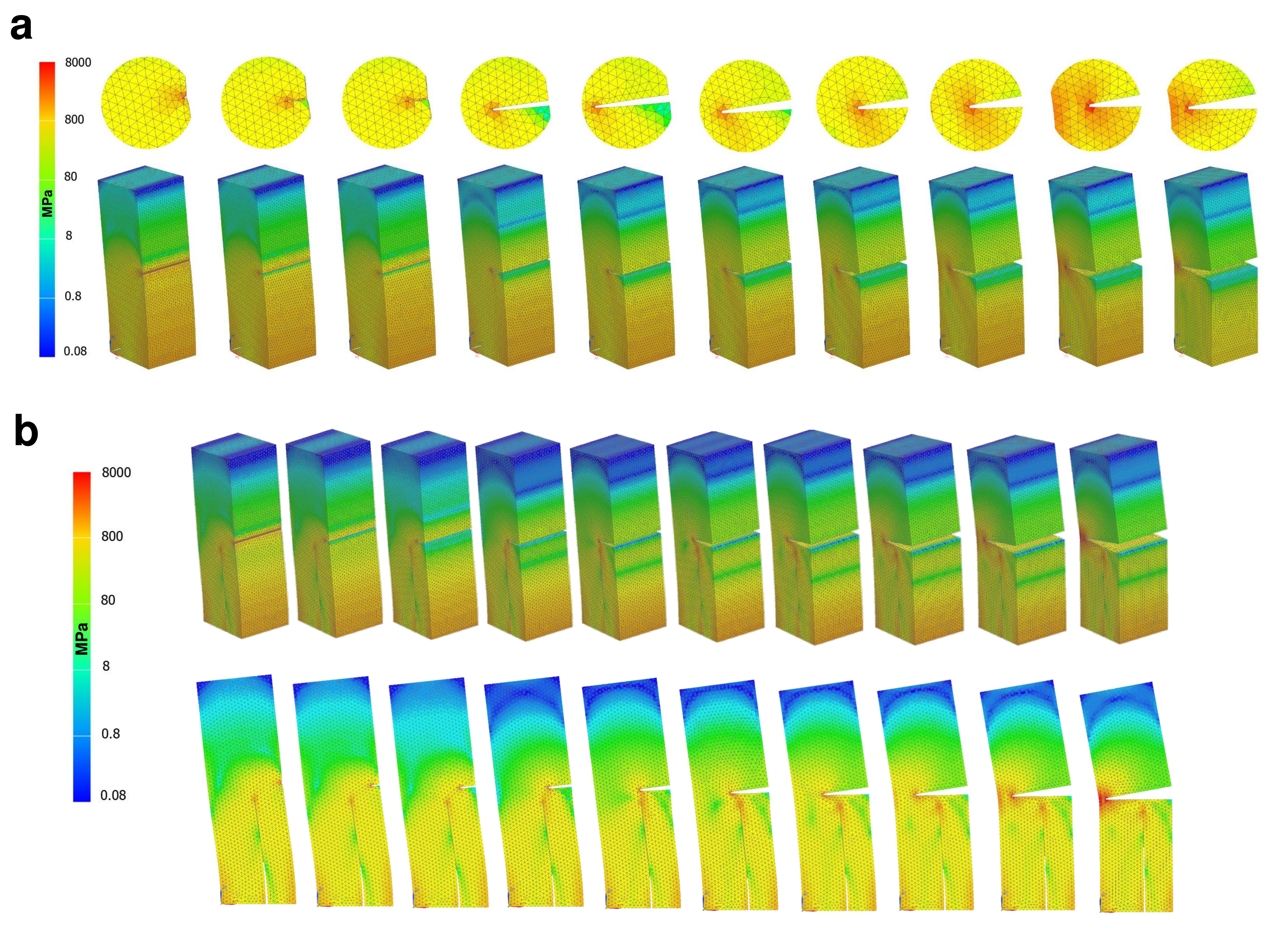}
	\end{center}
	\vskip -0.5 cm
	\caption{{\bf Simulations of the distribution of pressure in the sample during fracture.} Rectangular beam with a similar geometry as a URu$_2$Si$_2$ sample and a wedge that moves in each sequence (from left to right) through the sample due to a force applied to the top of the beam. The color scale represents the pressure distribution in MPa. The black lines are the mesh used in finite element calculations of the local pressure. The pressure is given by the color scale, or the bars on the left. The wedge opens from left to right, the latter image being just prior to fracture. {\bf a} Rupture of a sample made of a single piece, without defects. On the top row we show a view from the side of the fracture position for each image of the sequence. Notice that pressure is strongly built up on the wedge, which travels as a crack front through the surface. On the bottom row we visualize the pressure distribution all over the sample. {\bf b} Rupture of a sample with a wedge-like large defect that comes close to the sample cleaving plane. On the top row we show the pressure distribution all over the sample. On the bottom row we show the pressure distribution from the side all over the sample.}
	\label{Fig8_Pressure}
	\end{figure*}

In the sample cleaved a room temperature we found that it was not easy to identify the starting crack point. Instead we observe a starting crack zone of approximately a few $\mu m^2$ (red rectangle in Fig.\,\ref{Fig7_SEM}{\bf a} and Fig.\,\ref{Fig7_SEM}{\bf c}). By contrast, on the sample cleaved al low temperature, we identify the starting crack point as the red arrow in Fig.\,\ref{Fig7_SEM}{\bf d}. In every crack, we can identify the primary surface marks, called mirror and mist zones. These are regions where a crack radiates outwards from the starting point of fracture. These zones are the transition regions between the starting crack point and the hackle lines. We define the mirror zone by the region enclosed by the red dashed line (Fig.\,\ref{Fig7_SEM}{\bf d}). Inside the mirror zone we identify radially outgoing lines. Outside, in the mist zone, lines start to become straight and oriented with the crystalline axis, ending up in hackle lines. Hackle lines run in the local direction of craking and separate parallel but non coplanar sections of the fractured surface. Twist hackles are formed when the crack runs parallel to a preferent cleaving plane and the direction of the normal stress to this plane changes. Then, the crack cannot tilt in response to the new direction of stress direction and thus it splits into small, separated segments\cite{Quinn}. These lines form at exactly 90$^{\circ}$ to each other and are often close to being parallel to a cristalline axis. The lines evidence the large accumulation of strain before the crack (see also Fig.\ref{Fig8_Pressure}).

In order to address with more accuracy the crack features on the cleavage surfaces on both samples, we focus our attention in areas less than 25 $\mu m$ $\times$ 25 $\mu m$ (Fig.\,\ref{Fig7_SEM}{\bf e,f}). There we can observe important differences. In the sample cleaved at room temperature we identify flat surfaces of tens of microns with rectangular defects within areas of sizes 1-2 $\mu m^2$ that are parallel to the [10] and [01] directions (Fig.\ref{Fig7_SEM}{\bf e}). In the sample cleaved at 4.2K, by contrast, we observe stripes nearly parallel to the [01] atomic direction (Fig.\ref{Fig7_SEM}{\bf f}). These stripes are formed hackle lines that run close to a crystalline axis. These are related to the travelling direction of the crack front. The interaction of the crack front with dynamical modes of URu$_2$Si$_2$ defines the moir\'e one-dimensional charge modulation discussed in the main text.

The establishment of any permanent modification of the crystal structure requires energy. The only available source is the strain energy dissipating during cleavage, which is used to create the new surfaces obtained after cleaving. It is useful to discuss the strain we accumulate before breaking the sample. To see this, we can consider that, when we use samples of 1 mm square section (instead of 0.5 mm), we break a tin solder joint of the pulling mechanism instead of the sample. By decreasing sample thickness to 0.5 mm we break instead the sample. Thus, we can estimate that the uniaxial force on the sample is of order of the shear force of a tin solder joint. A good estimate is probably between 1 and 10 kg, say 5 kg or 50 N. This is applied as a shear to the sample using a wedge. The uniaxial pressure is as high as 200 kg/mm$^2$, if we estimate that contact between the wedge and the sample is along an area of 0.1x0.5 mm$^2$. This produces as much as 20 kbar uniaxial pressure, enough to locally drive the system into an ordered phase close to the wedge\cite{Mydosh11,Mydosh_2020}. More detailed calculations are provided in Fig.\,\ref{Fig8_Pressure}. We use a finite element calculation with the software NX Nastran. We use isoparametric tetrahedron elements with four vortex nodes and size additional midside nodes. The sample size is 1 mm $\times$ 1 mm $\times$ 3 mm and the mesh size is 50$\mu m$. We apply a lateral force of 50 N and a vertical tear force of 100 N, to be able to consider a sample separated in two parts in the model. We take a Young modulus of 192.85 GPa\cite{Wang_2020}. We see that we can easily obtain locally pressures up to 80 kbar (8 GPa) at the wedge (Fig.\,\ref{Fig8_Pressure}a). We also see that, in presence of a large defect inside the sample, the local pressure is considerably enhanced and modified at the tip of the defect (Fig.\,\ref{Fig8_Pressure}b). This suggests that large sample anomalies, like cuts or twists, which remain unnoticed in usual characterization experiments such as resistivity or specific heat, strongly increase the likelihood of inducing permanent modifications during fracture.

When the sample cleaves, the pressure is of course released. However, there is a large amount of available energy from the accumulated strain. We can estimate the released energy and compare it to the bonding energy and find very large values, of order of the mJ. On the other hand, usually the tensile strength upon shear of the material is exactly the bond-breaking strength, plus the energy required to produce surface modifications. To have such a surplus of mechanical energy that does not go into bond-breaking and creates permanent features on the surface, we need that the lateral edge of the sample has a higher tensile strength than the bulk. This should indeed be the case, taking into account that the sample has been cut and has certainly many more defects close to the edge. As we see in Fig.\,\ref{Fig7_SEM}, the surfaces obtained from cleaving an as-grown sample at room temperature are very different than those obtained cleaving a sample shaped using a wire saw at low temperature. The latter shows many more small scale modifications of the surface. Thus, the crack process has enough energy to produce permanent modifications of the surface. Furthermore, the local accumulated pressure drives the sample out of the HO state for the duration of the propagation of the crack front through the sample.

\subsection{Steps and the one-dimensional modulation}

\begin{figure*}
	\begin{center}
	\centering
	\includegraphics[width=0.9\textwidth]{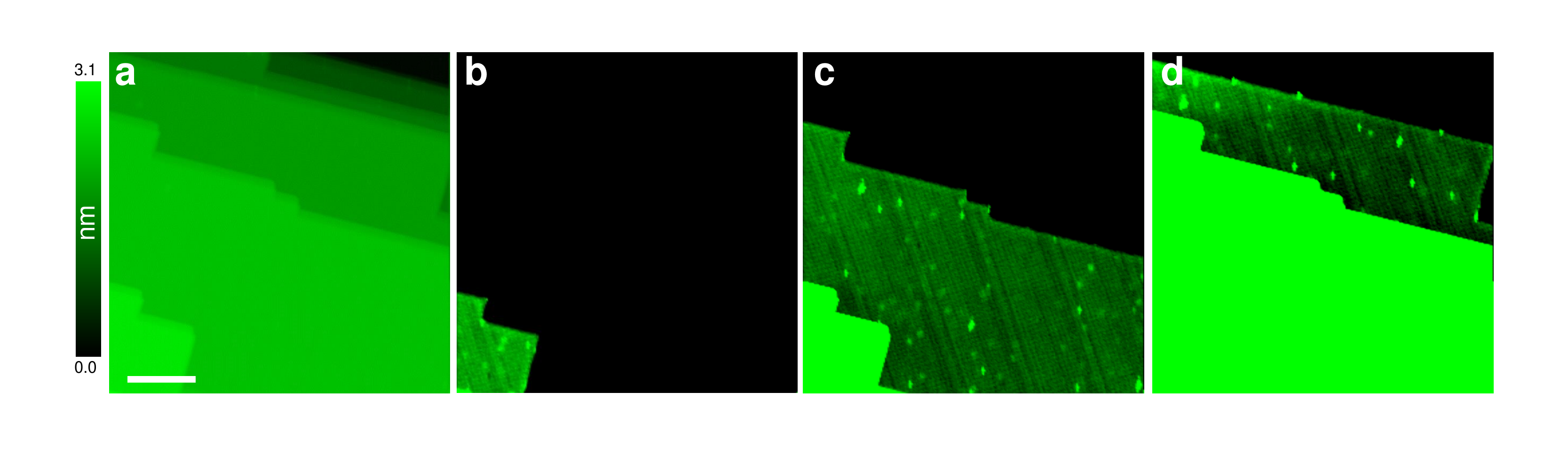}
	\end{center}
	\vskip -0.5 cm
	\caption{{\bf Moir\'e one-dimensional CDW and atomic size steps in URu$_2$Si$_2$.} {\bf a} Field of view with different atomic size steps. The white bar is 23 nm in size. Current is held constant at 4 nA and bias voltage is of 2 mV. The bar at the left provides the height in nm as a color scale. Notice that there is a structure of well defined steps, separated by steps, each one being half the c-axis unit cell size. In {\bf b,c,d} we modify this color scale to provide maximum contrast at three plateaus between steps. Notice that we can clearly observe, in addition to the crystal lattice, the one-dimensional modulation which goes from top left to bottom right.}
	\label{Fig9_Steps}
	\end{figure*}

By contrast to the large steps observed using optical microscopy and SEM, it is interesting to analyze the one-dimensional modulation in atomically sized steps with STM. In Fig.\ref{Fig9_Steps} we show a field of view with three large steps, each one consisting of a change in height of exactly half of the c-axis lattice constant. In the resulting plateaus, which are atomically flat, we can clearly see the modulation when we increase the contrast in the color scale that shows the topography (Fig.\ref{Fig9_Steps}{\bf b,c,d}).

\subsection{Displacement vector}

\begin{figure}
	\begin{center}
	\centering
	\includegraphics[width=0.95\textwidth]{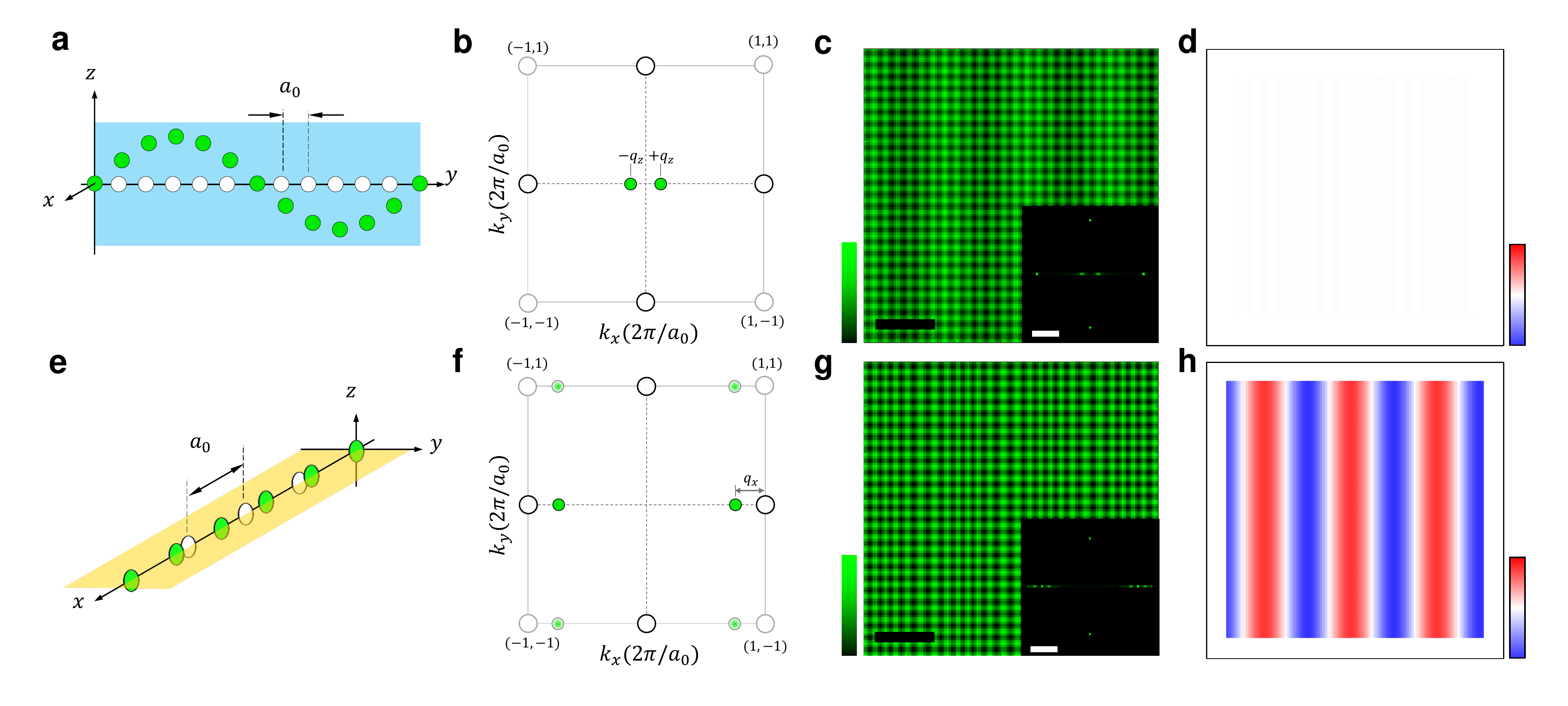}
	\end{center}
	\vskip -0.5 cm
	\caption{{\bf Displacement vector in real and reciprocal space.} Schematic representation of a real space modulation for atomic displacements in a square attomic lattice. White circles show atomic positions without perturbation and green circles with a sinusoidal perturbation. In {\bf a-d} the perturbation is along the c-axis, out of plane. In {\bf e-h} we show a longitudinal modulation along the x-axis. The positions in reciprocal space of the Bragg peaks in the Fourier transforms are shown in {\bf b,f} and in the insets of the corresponding images {\bf c,g}. White dots in {\bf b,f} are the atomic Bragg peaks and green dots the peaks due to the modulation. Notice that peaks appear at the center of the Fourier transform only when there is an out of plane modulation. In {\bf d,h} we show the associated strain map, obtained as discussed in the text. Black scale bars are five times the lattice constant long. The color scale goes from zero (black) to one (green) in {\bf c,g} and from blue $-$10\% to red $+$10\% of the lattice constant in {\bf d,h}.}
	\label{Fig10_Vector}
	\end{figure}

As we show in Fig.\,\ref{Fig10_Vector}, the Fourier patterns of displacement vectors with different directions are different. We take $\boldsymbol{D(\boldsymbol{r})}=\boldsymbol{A}cos(\boldsymbol{q}\boldsymbol{r})$, with wavevector $\boldsymbol{q}$. In Fig.\,\ref{Fig10_Vector}{\bf a,c} we show a modulation that occurs along the c-axis, out of plane ($\boldsymbol{A}=(0,0,A)$). In the Fourier transform Fig.\,\ref{Fig10_Vector}{\bf b} and inset of In Fig.\,\ref{Fig10_Vector}{\bf c} we observe a set of peaks around the center. When there is an in-plane displacement (Fig.\,\ref{Fig10_Vector}{\bf e,g}), the peaks in Fourier space appear around the Bragg peaks of the unmodulated atomic lattice (Fig.\,\ref{Fig10_Vector}{\bf f} and inset of {\bf g}), without peaks at the center of the Fourier transform.

In our images of the one-dimensional modulation, we observe peaks at the center of the Fourier transform. Thus, we have a charge modulation exclusively along the c-axis, in the same way as the modulation $\boldsymbol{A}=(0,0,A)$ mentioned above (Fig.\,\ref{Fig10_Vector}{\bf a-d}). Sometimes we also observe satellite modulation peaks around atomic Bragg peaks but with smaller amplitude that the peaks appearing at the center of the Fourier transform.

It is interesting to analyze the possibility of having lateral atomic displacements. To this end, let us start by remarking that when $\boldsymbol{A}$ has in-plane components, atoms are displaced with respect to their equilibrium positions within the plane. We can thus define an in-plane field of displacements $\boldsymbol{u}(\boldsymbol{r})$ and describe the same modulation as displacements with respect to an undisturbed perfect lattice. If the wavevector of the modulation is sufficiently separated from the wavevector of the atomic lattice, we can obtain the displacements and the strain map $M$, which is $M(\boldsymbol{r})\propto\frac{\partial u_x(\boldsymbol{r})}{\partial x}+\frac{\partial u_y(\boldsymbol{r})}{\partial y}$. The strain map $M$ can be normalized to the atomic lattice constant $a$ and shows quantitavely how much the atomic separation is reduced or extended in presence of a modulation or a defect. Let us see this in a couple of examples.

First, in Fig.\,\ref{Fig10_Vector}{\bf a} there is no in-plane strain (Fig.\,\ref{Fig10_Vector}{\bf d}), so that $M$ is equal to zero and we have a perfect square lattice. In a longitudinal modulation, by contrast, Fig.\ref{Fig10_Vector}{\bf e}, there are in-plane atomic displacements and thus local strain (Fig.\ref{Fig10_Vector}{\bf h}). The Fourier transform shows peaks indicating the direction of the modulation (here one-dimensional) surrounding the Bragg peaks of the atomic lattice, see Fig.\ref{Fig10_Vector}{\bf f}. Let us now see how we calculate the strain maps $M(\boldsymbol{r})$ of \ref{Fig10_Vector}{\bf d,h}.

The strain map $M(\boldsymbol{r})$ can be obtained by operating on an image showing a lattice with strain. We define the atomic lattice as a periodic cosinusoidal modulation $A(\boldsymbol{r})=Fcos(\boldsymbol{k}\boldsymbol{r-u(r)})$ with amplitude $F$ and wavevector $\boldsymbol{k}$. Notice that we calculate the strain from a local displacement vector $\boldsymbol{u(r)}$. To obtain $\boldsymbol{u(r)}$ we calculate $A'(\boldsymbol{r})=\sum_{\boldsymbol{r'}}A(\boldsymbol{r'}) \frac{1}{2\pi L^2}e^{-(\boldsymbol{r}-\boldsymbol{r'})^2/2L^2}$. We chose $L$ in such a way as to be able to capture strain whose spatial variation is of order of $L$. Of course, $L$ should be much larger than interatomic spacing $a$ and much smaller than the field of view. Then, $A(\boldsymbol{r})\propto cos(\boldsymbol{k}\boldsymbol{u(r)})$, because the Gaussian weighting ($\frac{1}{2\pi L^2}e^{-(\boldsymbol{r}-\boldsymbol{r'})^2/2L^2}$) occurs over a lengths scale that involve many interatomic distances, which eliminates the small wavelength contributions. We can then invert the cosine, taking care to correct for phase slips, and obtain $\boldsymbol{u(r)}$. From $\boldsymbol{u(r)}$ we can reconstruct the strain free lattice and calculate the strain map $M(\boldsymbol{r})$ (Fig.\ref{Fig10_Vector}{\bf d,h}). This is a widely used algorith with many applications, including correction of drifts and measurement of strain in strained lattices\cite{Lawler2010,Zeljkovic2015}. In Fourier space, it corresponds to extract the Fourier components around the atomic lattice Bragg peaks along x and along y and shift these together down to the center of Fourier space. The result is that a strain map $M(\boldsymbol{r})$ is visualized as a two-dimensional image, as shown for the above mentioned examples in \ref{Fig10_Vector}{\bf d,h}. Notice that $M(\boldsymbol{r})$ can be calculated only at a distance $L/2$ from the border of the image.

\begin{figure*}
	\begin{center}
	\centering
	\includegraphics[width=0.8\textwidth]{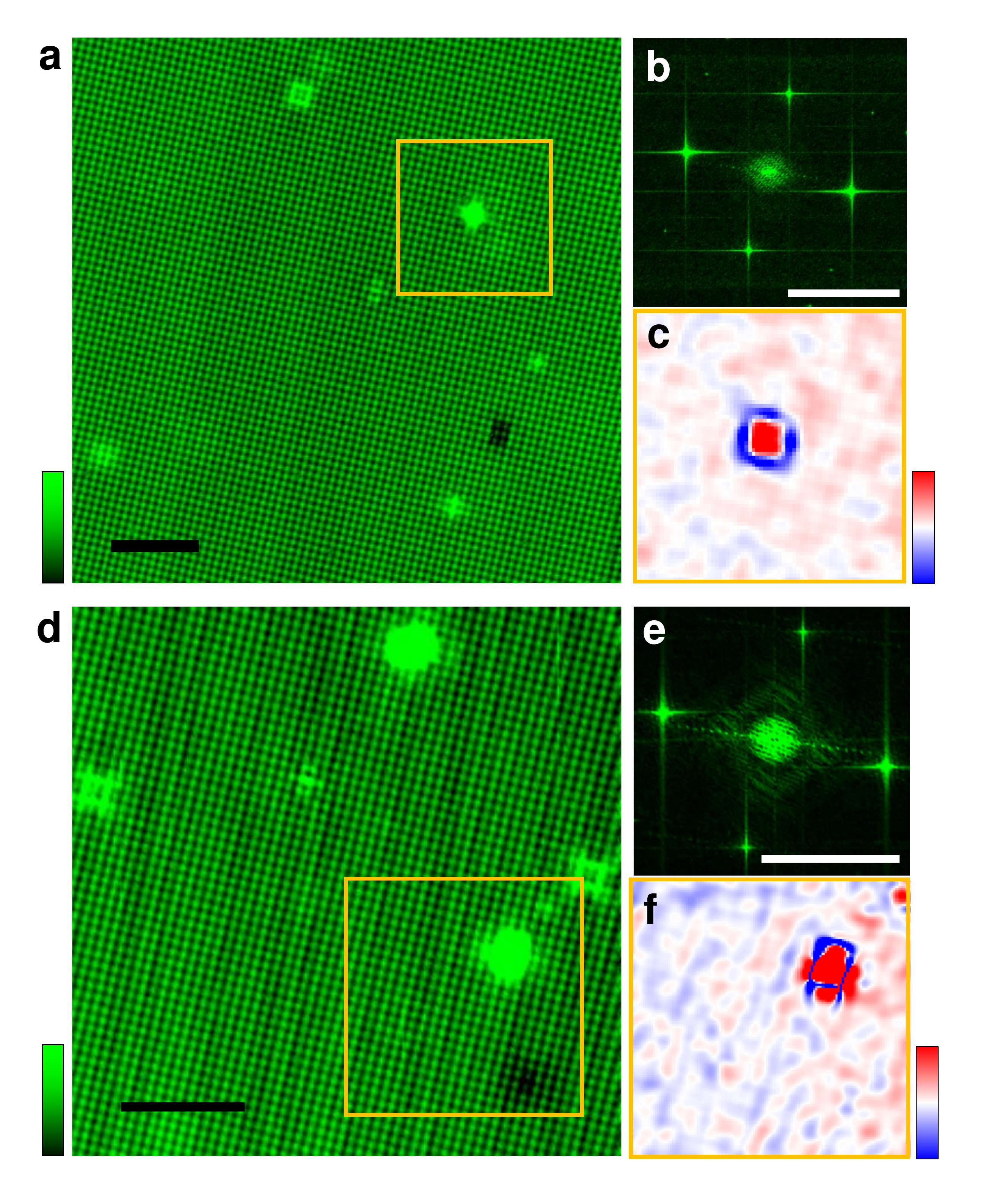}
	\end{center}
	\vskip -0.5 cm
	\caption{{\bf In-plane strain.} In {\bf a} we show a topographic image (made with a tunneling current of 0.2 nA and a bias voltage of 9 mV). The black scale bar is 5 nm long. The corresponding Fourier transform is shown in {\bf b} (white scale bar is 3 nm$^{-1}$ long). In {\bf c} we show the in-plane strain map of the orange square shown in {\bf a}. In {\bf d} we show the topography of another field of view (made with a tunneling current of 0.5 nA and a bias voltage of 4 mV), in {\bf e} its Fourier transform and in {\bf f} the strain on the area marked with an orange square shown in {\bf d}. The color scale of the topographic images is shown on the bottom left and goes from 0 pm (black) to 20 pm (green). The color scale of the strain is shown on the bottom right and ranges $\pm 5$\% of the in-plane lattice constant $a$. The strain maps can be calculated at a distance from the border of order of $L/2$. We choose $L$ in such a way as to be able to identify strain related to the modulation, between 3 and 4 $a$.}
	\label{Fig11_Strain}
	\end{figure*}

We can now apply such a reasoning to our data. As we discuss in the paper, in most cases the Bragg peaks associated to the modulation are centered at the origin. We rarely see a signal around the Bragg peaks and it is much weaker than the signal at the center. Therefore, the modulation has only a c-axis component and we indeed do not observe any strain. To test this with more detail we proceeded to the calculations and show the result in Fig.\,\ref{Fig11_Strain}. Let us start with an image with a weak modulation (smaller corrugation) Fig.\,\ref{Fig11_Strain}{\bf a}. The one-dimensional modulation is seen in the Fourier transform Fig.\,\ref{Fig11_Strain}{\bf b} as a set of peaks at the center, with no clear signature around the Bragg peaks of the atomic lattice. The corresponding strain map $M(\boldsymbol{r})$ is flat within our noise level, which we estimate to be of 0.1\% the lattice constant $a$. For comparison, the vertical displacement we can associate to the one-dimensional modulation is at most of 2 pm, which is 0.5\% of the lattice constant $a$. A pressure of 10 kbar corresponds to a modification of the lattice constant by about 1 pm.

Interestingly, we do observe an increase in the displacement around atomic size defects. We show $M(\boldsymbol{r})$  around such a defect in Fig.\,\ref{Fig11_Strain}{\bf c}. The defect consists of a square of Si atoms on top of the U surface. The strain field is a square which follows the orientation of the Si atoms (at 45$^{\circ}$ with respect to the U lattice). It shows that the strain accumulates in the U lattice just around the defects. The interatomic distance between Si atoms is also modified. Let us now consider a field of view where the one-dimensional modulation has a slightly larger corrugation (Fig.\,\ref{Fig11_Strain}{\bf d}). The Fourier transform also shows mostly peaks around the center (Fig.\,\ref{Fig11_Strain}{\bf e}). When calculating $M(\boldsymbol{r})$ we find a slight tendency to form one-dimensional stripe. Around a defect (\ref{Fig11_Strain}{\bf f}) we identify here an asymmetry in the strain map around the defect. The compression and extension occurs preferentially along an axis that is parallel to the one-dimensional modulation. This shows that there is an interaction between the one-dimensional modulation and the strain field produced by individual defects.

\subsection{Bias voltage dependence of STM topographic images and spectroscopic features of the one-dimensional modulation}

\begin{figure*}
	\begin{center}
	\centering
	\includegraphics[width=0.9\textwidth]{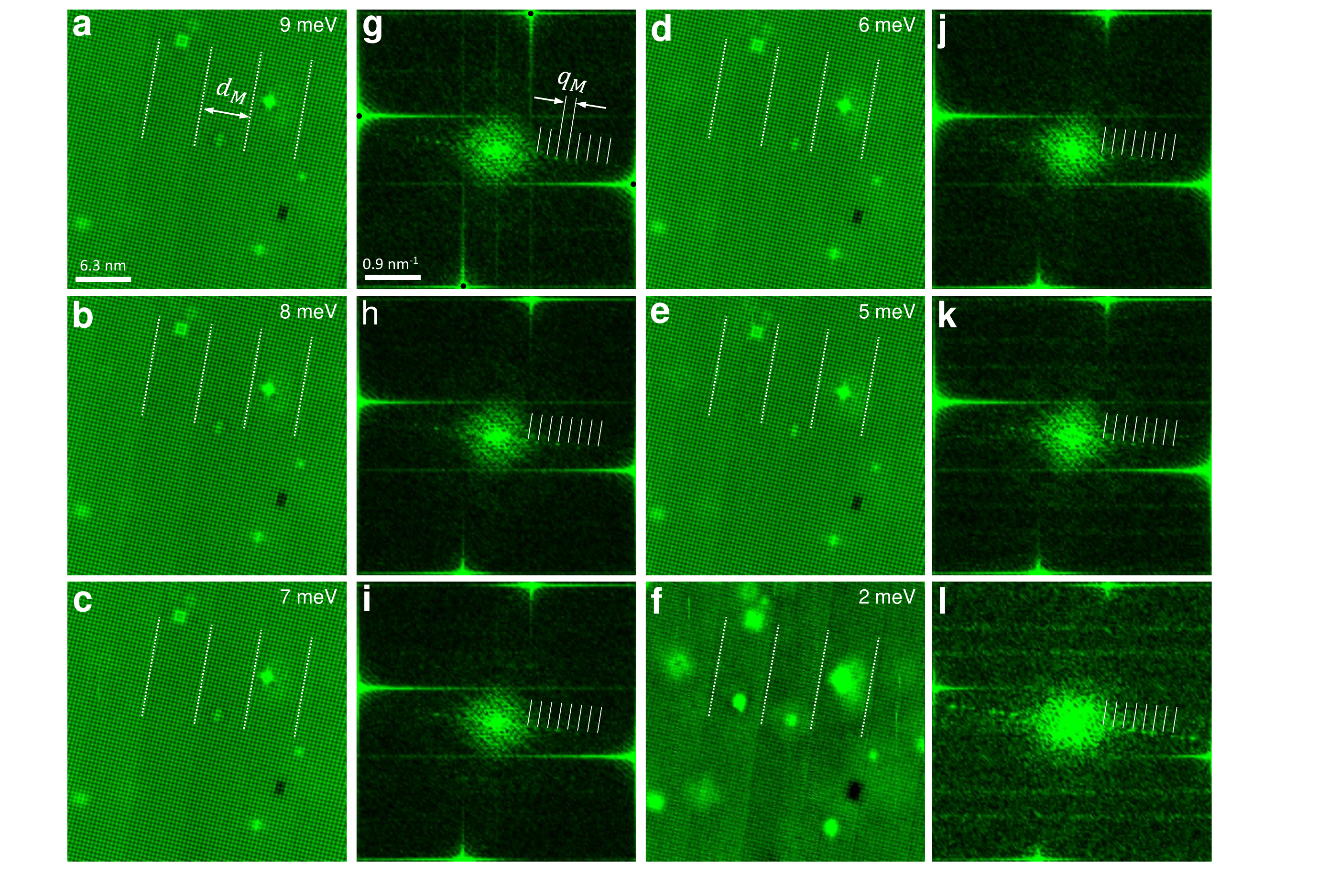}
	\end{center}
	\vskip -0.5 cm
	\caption{{\bf Bias voltage dependence of the 1D-CDW.} {\bf a} shows atomically flat images made at the bias voltage marked on the top right part of the image. We mark the modulation at $d_M$ by white dashed lines. Scale bar is on the bottom left. Current is held constant at 0.2 nA. In {\bf g} we show a zoom of the corresponding Fourier transform around its center, with the scale bar on the bottom left. We highlight the Bragg peaks of the crystal lattice by dark points and by white dashed lines the modulation wavevector $q_M$. This is repeated for different bias voltages in the other panels ({\bf b,h} to {\bf f,l}). The color scale in the real space images spans 28 pm and in Fourier space 2 pm  from black to green.}
	\label{Fig12_Bias}
	\end{figure*}

In Fig.\,\ref{Fig12_Bias} we show the bias voltage dependence of the modulation discussed in the main text. We focus here on topographic images taken in the range between 10 mV and 2 mV. In real space we see the one-dimensional modulation and a few atomic size defects. The Fourier transform shows the four Bragg peaks of the atomic lattice and the peaks corresponding to the $q_M$ modulation. In Fourier space we also observe a white circle, with structure inside, around the center. This is the consequence of quasiparticle interference scattering at the defects. Here we show topographic images, which are obtained at constant current. The current is given by $I(V) \propto \int_0^{eV_B}dE N(E-eV)$ where $V_B$ is the bias voltage given in the images and $N(E)$ the density of states. Therefore, it provides an energy integrated account of the density of states between zero bias and the voltages mentioned at each image in Fig.\,\ref{Fig12_Bias}.

The fact that the observed corrugation is so small suggests that the associated features in $N(E)$ are very small too. Nevertheless, we can try to discuss the features in the spectroscopy, first remembering previous results \cite{Aynajian10, Schmidt10,Haule09} and comparing ours to those results and then analyzing in more detail in the low energy part of the data.

	\begin{figure*}
	\begin{center}
	\centering
	\includegraphics[width=0.9\textwidth]{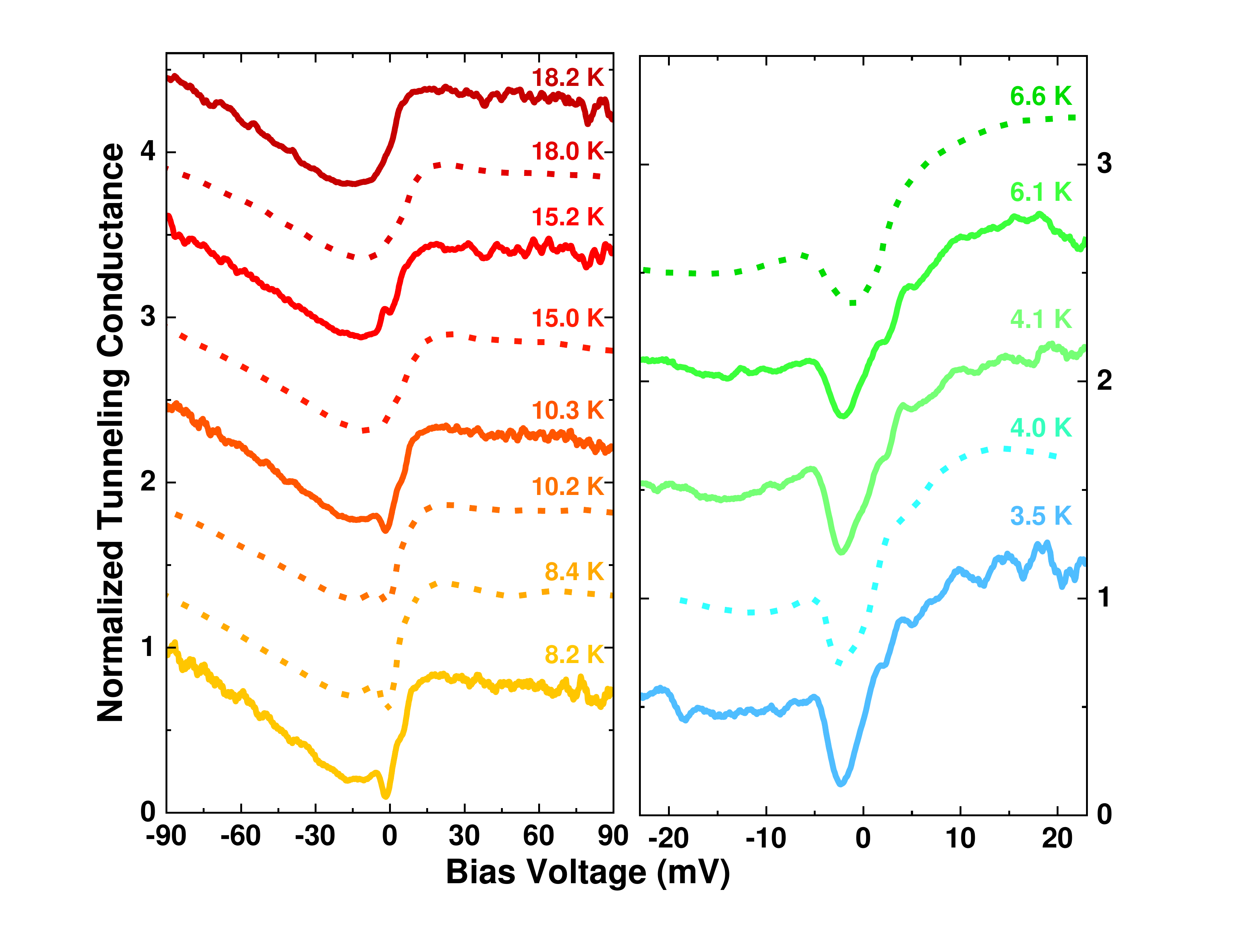}
	\end{center}
	\vskip -0.5 cm
	\caption{{\bf Comparison to previous work.} Tunneling conductance as a function of the bias voltage for different temperatures, provided in the figure. Notice that the bias voltage scale is up to 90 mV in the left panel whereas we focus on the low bias voltage part in the right panel, at low temperatures. In particular, the Fano lineshape between $\pm$ 30 mV seen in the left panel is just an asymmetric background in the right panel. Our data are the lines. Dashed lines are data from \cite{Aynajian10}.}
	\label{Fig13_Compare}
	\end{figure*}

We first have to realize that the local density of states has no region in energy where it is flat. It consists of a Fano anomaly, superposed to a gap opening and of two van Hove singularities located at a few mV. To see this, let us start by comparing our work with previous results (Fig.\ref{Fig13_Compare}). As shown in Refs.\cite{Aynajian10, Schmidt10,Haule09}, the local density of states in URu$_2$Si$_2$ has several interesting features whose evolution is strongly linked to the heavy fermion nature of URu$_2$Si$_2$ and the HO transition. When cooling from room temperature, the local density of states acquires an asymmetric shape. The asymmetric shape follows well a Fano-lineshape. The Fano-lineshape occurs due to interference from simultaneous tunneling into light and heavy bands. When cooling, the formation of the heavy bands is seen in the establishment of the Fano-lineshape in the local density of states. At about 18 K (see Fig.\ref{Fig13_Compare}), the resulting local density of states is explained by well established heavy electron bands \cite{Aynajian10, Schmidt10,Haule09}. When entering the HO state, a feature appears close to zero energy. The feature develops when cooling into a V-shaped depression of the local density of states, which is asymmetric in bias voltage (see Fig.\ref{Fig13_Compare}) \cite{Aynajian10, Schmidt10,Morr_2016}. As we see in Fig.\ref{Fig13_Compare}, we can exactly reproduce those previous results.

	\begin{figure*}
	\begin{center}
	\centering
	\includegraphics[width=0.85\textwidth]{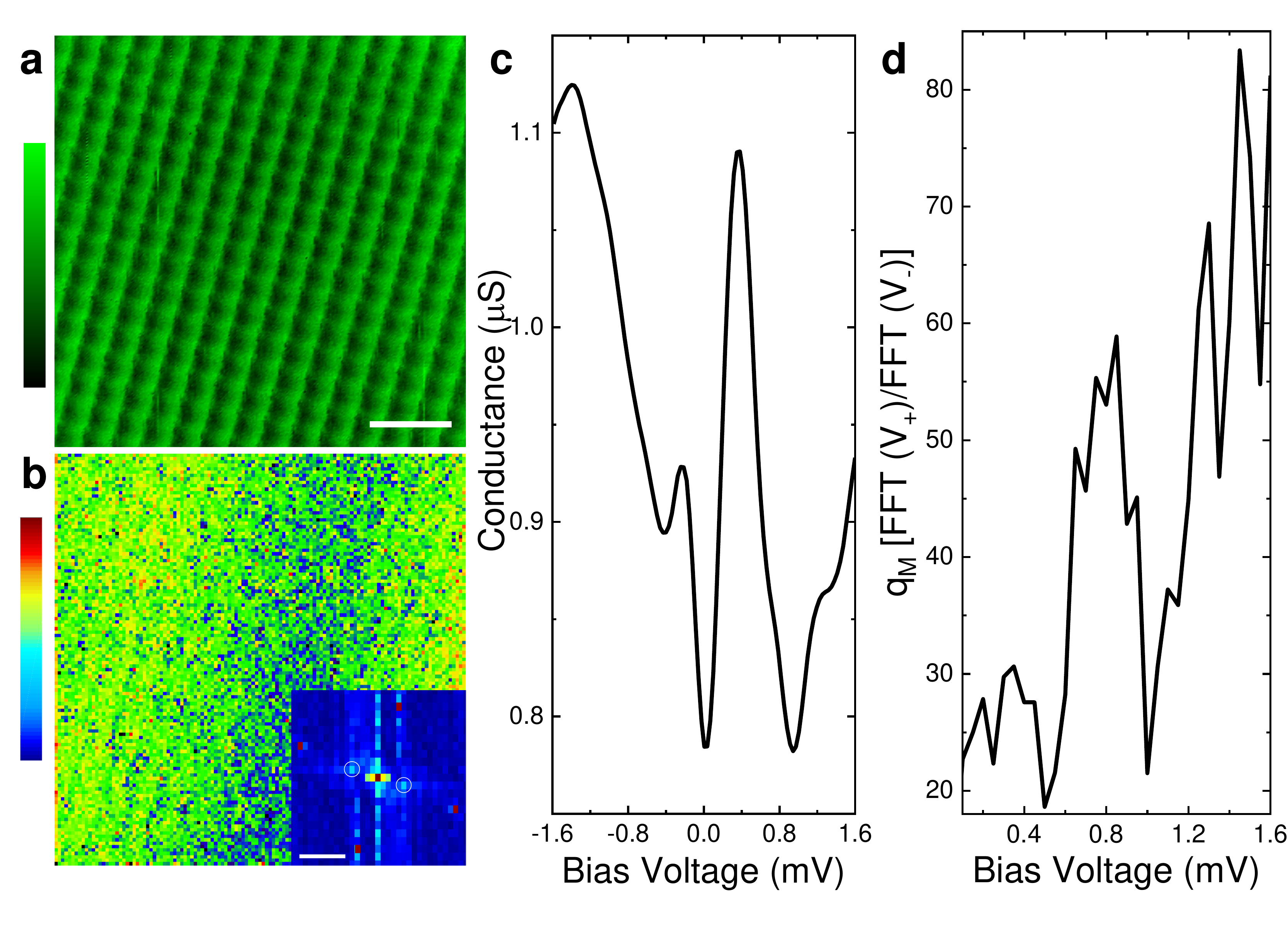}
	\end{center}
	\vskip -0.5 cm
	\caption{{\bf Topography and spectroscopy at very low bias voltages and temperatures.} In {\bf a} we show a topography in a small field of view free of defects. White scale bar is 1.3 nm long. In {\bf b} we show the map of the tunneling conductance at a bias voltage of 0.85 mV divided by the map at -0.85 mV. The corresponding Fourier transform is shown in the lower right inset, with the white scale bar being 1.4 nm$^{-1}$. White circles provide the peaks of the one-dimensional modulation. Atomic Bragg peaks are in red. In {\bf c} we show the average tunneling conductance curve over the whole image. In {\bf d} we show the amplitude of the Fourier peak marked by white circles in the inset of {\bf b}.}
	\label{Fig14_Atomic}
	\end{figure*}

As shown in Refs.\cite{Aynajian10,Schmidt10,Morr_2016}, the feature reflects the modifications in the bandstructure introduced by HO, observed thanks to enhanced quasiparticle interference induced by Th in doped samples\cite{Schmidt10}. These can be associated to a gap opening due to nesting along $q_0$ when entering HO that provides the V-shaped depression. It was also previously shown that the establishment of the Fano-lineshape leads to peaks from the band hybridization in the local density of states. These are due to van Hove anomalies at the energies with strongest band bending that provide a large density of states \cite{PhysRevLett.103.206402}. These occur at particularly low energies in URu$_2$Si$_2$\cite{Morr_2016}. In addition, there are two hybridized bands\cite{Morr_2016,PhysRevB.86.035129}. These hybridization features occur below 2mV in form of several kinks in the bandstructure. Th doped samples show enhanced in-plane electronic scattering along a few wavevectors, some close to the $q_1$ wavevector providing a hot spot on the Fermi surface\cite{Schmidt10}.

It should be clear from the outset that any feature appearing in such a wavy background is very difficult to disentangle, particularly if associated to such a small signal in the topography. Below about 2 mV, we can identify some features in the spectroscopic maps. These are very weak and yet not fully clear, although worth to comment on.

Some of these features are seen when zooming into data obtained in Ref.\cite{Aynajian10,Schmidt10} but these were mostly left undiscussed. Further analysis shown in Ref.\cite{PhysRevLett.103.206402} related these features to the bandstructure. There are hybridized bands at very low energies with a bandstructure that is highly asymmetric with respect to the Fermi level\cite{Morr_2016}.  We show our result in \ref{Fig14_Atomic} on a field of view without defects. There are two clear peaks located at about -1.5 mV and at +0.4 mV. These peaks are at a a similar energy position as those found in previous work and related to the bandstructure. Here we observe that these are considerably more pronounced. As band hybridization leads to a highly asymmetric local density of states, we have chosen to divide the signal at positive with that at negative bias voltages. We then observe the modulation as a stripe which goes from the top left to the upper right part of the image. The Fourier transform shows the corresponding peaks, which show no dispersion with bias voltage, although their amplitude strongly varies. We observe that the amplitude decreases below about 1.6 mV. Furthermore, we observe an increase at about 0.8 mV (Fig.\,\ref{Fig14_Atomic}{\bf d}). This can highlight a modification of the signal related to the one-dimensional modulation at a van Hove anomaly. Of course, a comprehensive understanding of all these features requires a careful analysis of the tunneling conductance and its temperature dependence, but this already establishes a link between the one-dimensional modulation and the local density of states.

\bibliographystyle{apsrev4-1-titles}

%

\end{document}